\documentclass[sigconf]{acmart}

\usepackage{booktabs} 
\usepackage{eurosym}
\usepackage{subcaption}

\copyrightyear{2018} 
\acmYear{2018} 
\setcopyright{acmlicensed}
\acmConference[e-Energy '18]{The Ninth International Conference on Future Energy Systems}{June 12--15, 2018}{Karlsruhe, Germany}
\acmBooktitle{e-Energy '18: The Ninth International Conference on Future Energy Systems, June 12--15, 2018, Karlsruhe, Germany}
\acmPrice{15.00}
\acmDOI{10.1145/3208903.3208937}
\acmISBN{978-1-4503-5767-8/18/06}

\begin{document}
\title[Augmenting DER through local energy markets and dynamic phase switching]{Augmenting DER hosting capacity of distribution grids through local energy markets and dynamic phase switching}

\author{Jos\'e Horta}
\affiliation{%
  \institution{Telecom Paristech}
  \streetaddress{23 Avenue d'Italie}
  \city{Paris} 
  \state{France} 
  \postcode{75013}
}
\email{jose.horta@telecom-paristech.fr}

\author{Daniel Kofman}
\affiliation{%
  \institution{Telecom Paristech}
  \streetaddress{23 Avenue d'Italie}
  \city{Paris} 
  \state{France} 
  \postcode{75013}
}
\email{daniel.kofman@telecom-paristech.fr}

\author{David Menga}
\affiliation{%
  \institution{EDF R\&D}
  \streetaddress{EDF Lab Paris-Saclay}
  \city{Palaiseau} 
  \country{France}}
\email{david.menga@edf.fr}

\author{Mathieu Caujolle}
\affiliation{%
  \institution{EDF R\&D}
  \streetaddress{EDF Lab Paris-Saclay}
  \city{Palaiseau} 
  \country{France}}
\email{mathieu.caujolle@edf.fr}


\begin{abstract}
The limited capacity of distribution grids for hosting renewable generation is one of the main challenges towards the energy transition. Local energy markets, enabling direct exchange of energy between prosumers, help to integrate the growing number of residential photovoltaic panels by scheduling flexible demand for balancing renewable energy locally. Nevertheless, existing scheduling mechanisms do not take into account the phases to which households are connected, increasing network unbalance and favoring bigger voltage rises/drops and higher losses. In this paper, we reduce network unbalance by leveraging market transactions information to dynamically allocate houses to phases using solid state switches. We propose cost effective mechanisms for the selection of households to switch and for their optimal allocation to phases. Using load flow analysis we show that only 6\% of houses in our case studies need to be equipped with dynamic switches to counteract the negative impact of local energy markets while maintaining all the benefits. Combining local energy markets and dynamic phase switching we improve both overall load balancing and network unbalance, effectively augmenting DER hosting capacity of distribution grids.
\end{abstract}

%
%



\maketitle

\section{Introduction}
Distributed renewable energies are one of the main enablers for the energy transition, but their potential is currently being hindered by the limited capacity of distribution grids for hosting renewable energy, due to their negative impacts on quality of supply (QoS), in terms of peak flows, voltage deviations and network unbalance.

Local energy markets have been recently proposed as a solution for integrating the increasing number of residential photovoltaic panels (PV). These markets incentive prosumers\footnote{Consumer evolution towards pro-active participation on grid activities.} to schedule their distributed energy resources\footnote{Flexible loads, controllable generation and storage resources.}  (DER) in a way that reduces peak flows through the transformer and corresponding losses \cite{horta2017}. However, as most demand side management mechanisms, the schedule provided by such markets does not take into account the phases to which households are connected to, creating further uneven distribution of flows across phases. Such network unbalance degradations favor larger voltage rises/drops, augment line losses \cite{Mendia}, and strongly affect the lifespan of three-phase loads. Network unbalance is measured using the Voltage Unbalance Factor (VUF), which has to be kept under 2\% as one of the main QoS metrics.

The purpose of this work is to propose and evaluate mechanisms for taking advantage of all local market benefits without suffering from the mentioned network unbalance issues. We propose to combine local markets with dynamic phase switching, reallocating households to phases every time the local market decides on the flows to be exchanged among households. In particular, we propose cost effective mechanisms for the selection of households to be switched and for their optimal allocation to phases. The solution is based on solid state switches (STS) installed at the Point of Common Coupling (PCC) of some of the market participants.

The article is organized as follows. First, in Section \ref{sec:SysDescPhSwitch}, we introduce the system under study. In Section \ref{sec:RelWorkPhSwitch}, we discuss related work. Then, in Section \ref{sec:ModelPhSwitch}, we introduce our dynamic phase switching mechanisms, and in Section \ref{sec:SimulResultsPhSwitch}, we analyze the aggregated impact of markets and dynamic phase switching on distribution grid QoS. Finally, we conclude the paper and share some perspectives.

\section{System description}
\label{sec:SysDescPhSwitch}
The low voltage distribution grid under study consists of one Medium Voltage/Low Voltage (MV/LV) transformer and one or more feeders to which households are connected. All houses have a smart meter\footnote{France is massively deploying smart meters and has recently introduced specific regulation with respect to auto-consumption at the neighborhood level \cite{Decret}.} and some of them are equipped with a PV panel and/or a battery. The flows from the PV and to/from the battery can be controlled by means of smart inverters \cite{pignier2013energy}, which enable houses to control the destination of their renewable energy as well as the source of the energy that satisfies load demands. Such control over flows enables households to exchange energy and services through local markets (Section \ref{sec:PhSwitch_Sys_LEM}), providing data for dynamic phase allocation (Section \ref{sec:SysDynPhaseSwitch}) using solid state switches (Section \ref{sec:solidState}).

\subsection{Local energy market}
\label{sec:PhSwitch_Sys_LEM}
Local energy markets enable households to agree on the exchange of energy blocks and flexibility services with each other, service providers and DSO. In particular, we focus on the market proposed in \cite{horta2017}, which aims to balance renewable energy with flexible demand. They propose an auction mechanism and price incentives for synchronizing buy and sell offers. This market provides hour-ahead commitments from prosumers on the expected flows to be exchanged with the grid during 10-minute time slots and includes rewards for the enforcement of commitments\footnote{In this paper we assume that market commitments are perfectly enforced.}.

\subsection{Dynamic Phase switching}
\label{sec:SysDynPhaseSwitch}
The traditional solution to network unbalance is to reallocate the phases of some single-phase houses based on their average load profiles over long time periods. This requires a technician to manually switch phase connections, implying high costs and no adaptability. For these reasons, such a static approach is only applied when the VUF is close to 2\%. The aim of dynamic phase switching is to counteract the negative effect of scheduling mechanisms and to augment the hosting capability of the grid, rather than reallocating phases only when VUF reaches critical values. The deployment of dynamic phase switches can be done progressively by selecting households following the mechanism proposed in Section \ref{subsec:houseChoice}. Switches are operated every 10-minutes\footnote{time scope due to constraints imposed by regulation on voltage deviations \cite{EN50160, EN50438}.} as described in Section \ref{sec:allocationMechanism}, based on the corresponding market results and allocation of phases\footnote{Note that with smart meters we have access to the phase connection information.}.

\subsection{Solid-state transfer switches}
\label{sec:solidState}
The proposed implementation for the transfer switches is composed of three Si-IGBT\footnote{Interlocked Insulated-gate Bipolar Transistor switches. Alternative: Sic-MOSFET \cite{Tanabe}}, one for each phase, three mechanical contactors and a controller implementing a logic interlock. The switches are chosen by their high operating frequencies (200 Hz), reducing the time duration of break-before-make operations and respecting QoS requirements \cite{Shahnia2014}. These design decisions ensure that loads will not be affected by the switching operations, short-circuits are avoided thanks to the interlocking and switches are protected from faults by relying on standard mechanical contactors. The control signals are sent through the advanced metering infrastructure and relied to the STS by their corresponding TIC \cite{TIC} or BlueTic \cite{BlueTIC}.

\section{Related Work}
\label{sec:RelWorkPhSwitch}
Most scheduling mechanisms do not take into account the phase to which households are connected to. An example of mechanisms that do consider phase allocations are those based on Unbalanced Three-Phase Optimal Power Flow \cite{OPF}. Nevertheless, these require complex computation, detailed knowledge on the grid structure and do not consider dynamic phase switching explicitly. With respect to previous proposals for dynamic phase switching \cite{Mendia, Boroujeni, Shahnia2014}, they do not consider renewable energies, which have a major impact on grid unbalance and voltage deviations, and they require additional measurements points \cite{Boroujeni} or households to periodically share information on the flows exchanged with the grid \cite{Shahnia2014}, even though they have no incentive to share truthful information. In our work a local energy exchange market provides the incentives for households to reveal the information truthfully. Furthermore, existing mechanisms rely on complex algorithms for an extensive search over all possible phase switches. Instead, we believe the most appropriate approach is to find simple algorithms that would achieve considerable reductions of VUF with a reduced set of phase switches and with minimal requirements of information.

\section{Dynamic Phase Switch Model}
\label{sec:ModelPhSwitch}
\subsection{Phase allocation}
Let $\mathcal{H}$ denote the set of $H$ households on a LV distribution grid, from which a subset $\mathcal{M}$ participates on the local market. The houses in $\mathcal{M}$ are eligible for installing a dynamic switch as part of the set $\mathcal{E} \subset \mathcal{M} \subset \mathcal{H}$ of size $E$. Let denote the houses with static phases by $\mathcal{N}$ of size $N = H - E$. A phase allocation can be seen as a bipartite graph formed by the set of households and the set of phases, with a boolean adjacency matrix $X \in \mathbb{B}^{3\times H}$ where its element $x_{i,j}$ is $1$ if the household $j$ is connected to the phase $i$, and $0$ otherwise.

\subsection{Choice of households to switch}
\label{subsec:houseChoice}
In this section we propose simple heuristics to choose the houses where to install dynamic phase switches. We propose mechanisms that rely only on information that is already available to DSOs.

\subsubsection{Mean Based (MB)}
\label{subsubsec:MB}
The houses selected by the MB mechanism are those for which a DSO running a static phase allocation would decide to switch phases, based on long run average loads.

\subsubsection{Highest Average Flow (HAF)}
\label{subsubsec:HAF}
For HAF we first pre-select houses with PV that are assigned to the phase with the highest voltage and households without PV that are assigned to other phases, preferably the one with the lowest voltage. This aims to reduce production on the phase with highest voltage and/or to reduce consumption on the phase with lowest voltage. Then, we choose the one/s with the highest average flow (injection and demand respectively), in order to minimize the number of phase transfers required to balance flows across phases. This strategy is useful for the initial switch deployments, but it can be refined for better fitting the voltage unbalance gap and to further improve the performance of the system.

\subsubsection{Hybrid}
\label{sec:Hybrid}
In the hybrid approach MB is used for a pre-selection and HAF for obtaining the final set of candidates.

\subsection{Dynamic phase allocation}
\label{sec:allocationMechanism}
We propose to minimize the negative impact of market participants, which are in fact the ones that have the highest impact on voltage unbalance due to their DER. This requires the flows imposed by market participants to be as balanced as possible across phases. Let $e$ be the aggregate of commitments of market participants for the time slot on the corresponding feeder, we aim to obtain a phase allocation of the set of switches $\mathcal{E}$ such that the flow imposed by the set of market participants $\mathcal{M}$ on each phase is as close to the average $e_m = \frac{e}{3}$ as possible. For this, a natural metric to minimize is the least square distance of the corresponding flows. The binary vector $x = [x^a,x^b,x^c]^T \in \mathbb{B}^{3M}$ represents the phase allocation of market participants such that $X = [{x^a}^T,{x^b}^T,{x^c}^T] \in \mathbb{B}^{M\times3}$ is the corresponding adjacency matrix of the phase allocation graph. Then, the problem to be solved is

\begin{align}
\label{eq:lsq_minMarquetImpact}
&min_{x}|| \overline{e} - {P}^Tx ||^2 \\
&\textrm{s.t.} \nonumber \\
&x^a + x^b + x^c = \textbf{1} \label{subeq:1}\\ 
&x^0x^T = N \label{subeq:2}\\
&x_{i} \in {0,1}, \forall i \in \{1, \dots , 3M\} \label{subeq:3}
\end{align}

where $\overline{e} = [e_m,e_m,e_m]^T \in \mathbb{R}^3$, and $P \in \mathbb{R}^{3M\times3}$ is a block diagonal matrix formed from the vector of prosumer commitments $p_c \in \mathbb{R}^{M}$, such that the product $P^Tx$ denotes the vector of prosumer aggregated flows on the 3 phases for a given phase allocation $x$. The constraint (\ref{subeq:1}) ensures only one allocation per house is given, where \textbf{1} denotes the ones vector of size $M$. The constraint on the Hamming distance between $x^0$ and $x$ in (\ref{subeq:2}) ensures that houses of the set $\mathcal{N}$ will keep their phase allocations, as $x^0$ represents the current phase allocation, but with zeros on the elements corresponding to $\mathcal{E}$. This problem can be classified as Mixed Integer Least Squares (Mixed Integer Quadratic Program) and expressed as follows

\begin{align}
	\label{eq:quad_minMarquetImpact}
	&\operatorname{min}_x x^TQx + f^Tx \\
	&\textrm{s.t.} \nonumber \\
	&(\ref{subeq:1}) - (\ref{subeq:3})\nonumber
\end{align}

where $Q$ and $f$ can be expressed in function of the parameters of the original problem (\ref{eq:lsq_minMarquetImpact}) as follows: $Q = P^TP$ and $f = -P{\overline{e}}^T$.

\section{Simulations and Results}
\label{sec:SimulResultsPhSwitch}
\subsection{System scenarios}
\label{sec:SystemScenarios}
The evaluation process has two parts. First, in Section \ref{sec:ImpactAnalysis}, we evaluate the performance of dynamic vs. static phase switching. For this, we consider a scenario with massive deployment of DER and the same amount of phase switches for static and dynamic approaches. In this case, we consider $H = 33$ households, $8$ without flexibility (approx $H/4$), $8$ with battery only (approx $H/4$) and $17$ with PV and battery (approx $H/2$).

Then, in Section \ref{sec:AdaptAnalysis}, we evaluate the adaptability of both static and dynamic approaches as we evolve towards massive deployment of DER, represented by four consecutive scenarios with 50 houses:
\begin{itemize}
	\item Senario A: 20\% storage and 30\% renewable.
	\item Senario B: 40\% storage and 40\% renewable.
	\item Senario C: 60\% storage and 50\% renewable.
	\item Senario D: 60\% storage and 80\% renewable.
\end{itemize} 

\subsection{Simulation parameters}
We rely on the Distribution Network Simulation Platform (DisNetSimPl) developed by EDF R\&D, which provides an interface to OpenDSS \cite{OpenDSS}. We use a slightly modified version of the electricity network model provided in \cite{horta2017} and the following parameters:

\paragraph*{Load profiles -}6 summer days load profiles from SMACH \cite{amouroux2013simulating}.
\paragraph*{Production profiles -}Equal synthetic production for all houses. 
\paragraph*{Batteries -}We consider ideal batteries of 6 kWh capacity.
\paragraph*{Electricity prices -}Two levels Time Of Use pricing: 15 c\geneuro/kWh from 12 am to 4 pm and 20 c\geneuro/kWh from 5 pm to 11 pm.

\subsection{Results and discussion}
\subsubsection{Impact analysis}
\label{sec:ImpactAnalysis}
First, we analyze the impact of phase transfers on the benefits of local energy market interactions with respect to energy passing through the transformer, the corresponding losses on the transformer and the reductions on peak load. The benefits of local energy market are maintained for the phase switching mechanisms tested. As expected, these metrics are barely impacted by phase switching, as all are aggregated measures of load and losses at the transformer level. It is not worth to show the graphs, as the effect on peak flows is barely visible or negligible.

\begin{figure}[htb!]
	\centering
	\includegraphics[width=0.49\textwidth]{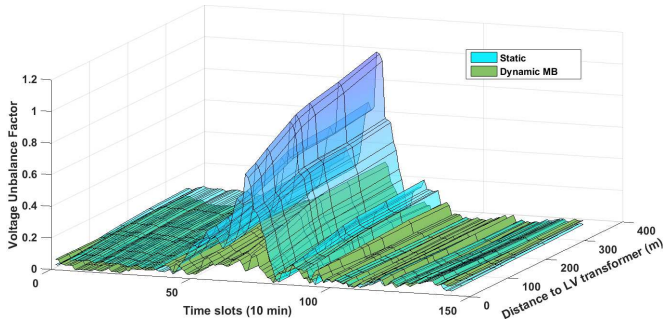}
	\caption{VUF variation along the highest loaded feeder.}
	\label{fig:VUF_surface}
\end{figure}

Then, we analyze the performance of phase transfer mechanisms on reducing maximum VUF and we illustrate the relevance of peak values of VUF in the presence of DER. In Figure \ref{fig:VUF_surface} we show the variation of VUF values over time for Day 1 along the highest loaded feeder. The comparison is made for four phase switches, with 3 being made (resp. installed) in the highest loaded feeder, following the static and Dynamic MB approaches. The figure illustrates the strong influence of renewable production and phase switching on voltage unbalance, up to the point that a couple of households with PV panels connected to the highest loaded phase can heavily degrade the quality of electricity supply. The current static approach fails to achieve consistent reductions of the peak VUF despite achieving reductions of mean VUF, while dynamic approaches perform much better for reducing peak VUF for the same amount of phase switches.

\subsubsection{Adaptability analysis}
\label{sec:AdaptAnalysis}
Next we analyze the adaptability of our dynamic phase switching solution compared to static switching along scenarios that represent a possible development of a residential distribution grid in the years to come, as introduced in Section \ref{sec:SystemScenarios}. First, we compare the amount of phase switches needed to achieve a reduction of peak VUF with respect to a scenario without market. Then, we determine the amount of dynamic phase switches in order to achieve, in all cases, a superior performance with respect to both VUF and Voltage rise/drop. 

\begin{figure}[htb!]
	\centering
	\includegraphics[width=\columnwidth]{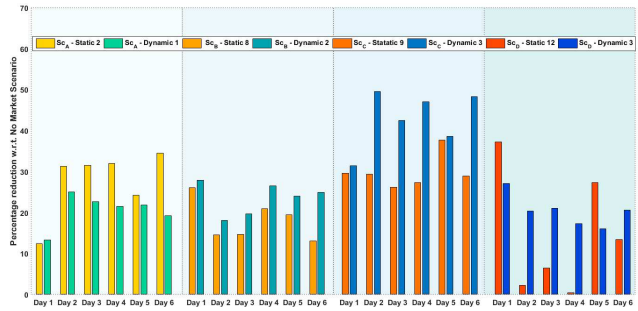}
	\caption{VUF reductions for 6\% dynamic phase switches.}
	\label{fig:VUF_scenariosAtoD_GoodEnough}
\end{figure}

In Figure \ref{fig:VUF_scenariosAtoD_GoodEnough} we observe the reduction of peak VUF that is achieved with the minimum number of phase switches needed to counteract the negative effect of the scheduling mechanisms for all days along all scenarios. The scenarios are represented as areas ordered from A to D, yellow to red colors represent the static approach, while green to blue represent dynamic results. For the static approach, 12 phase transfers are needed (2 in scenario A, 6 in B, 1 in C and 3 in D), while for the dynamic approach only 3 are enough (1 in A, B and C), which represents 24\% of households against only 6\% for the dynamic case. Note that for Scenario D, the static approach barely achieves the goal, which means that probably an extra transfer would be needed for different load conditions.

\begin{figure}[htb!]
	\centering
	\includegraphics[width=\columnwidth]{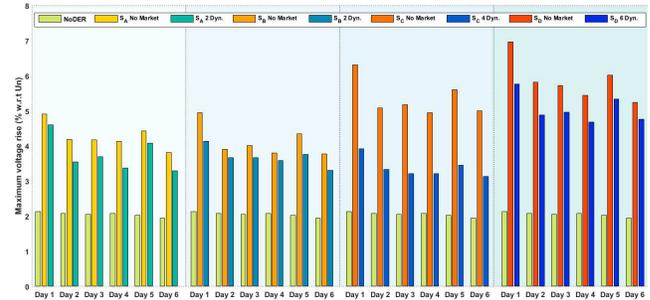}
	\caption{Voltage drop/rise for 12\% dynamic phase switches.}
	\label{fig:Vrise_scenariosAtoD_NoDER}
\end{figure}

If we increase the dynamic phase switches installed to 6 (2 in A, 2 in C, and 2 in D), representing 12\% of total households, we consistently achieve a superior performance. In Figure \ref{fig:Vrise_scenariosAtoD_NoDER} we show the voltage levels obtained in comparison with a scenario without market and a scenario without DER. We can see the voltage levels obtained at scenario D (80\% renewables) with our system are lower than those obtained in scenario C (50\% renewables) for a scenario without market. The same is valid for scenario C with respect to scenarios A and B. This represents a considerable increase in hosting capacity with respect to voltage rise/drop. With respect to VUF, we obtain for scenarios B, C and D, a similar performance to the scenario without DER, and that only requires 12\% of households to be equipped with dynamic phase switches.

Finally, with respect to losses, dynamic phase switching outperforms the static approach with less than half the deployments of technicians. The economic benefits obtained therein could be enough to finance the deployment of dynamic phase switches. The average total line losses reduction obtained across scenarios is of around $2.5 MWh$ per year, which would represent a reduction of $100 \geneuro/year$ for each 50 household neighborhood\footnote{This is not negligible as in France there are more than 30 million residential clients.}, considering an average electricity price of $40 \geneuro/MWh$. If we consider each scenario will last 5 years, during the first scenario the DSO will install 2 phase switches and recover $500 \geneuro$ in loss reduction, then during the scenario B there is no need for an additional switch, but around $500 \geneuro$ in losses would be saved. At the end of the 20 year period, for each 50 household neighborhood, the operator would have installed 6 switches and saved $2000 \geneuro$ in losses, which would mean more than $300 \geneuro$ for the deployment of each device. This is without taking into account the economic benefits of reducing VUF and Voltage rise/drop under massive deployment of DER. Such a discussion is only for illustrative purposes of the possible economic feasibility of dynamic phase switching\footnote{Note that OPEX costs are not considered in this analysis.}.

\section{Conclusions and perspectives}

A massive deployment of distributed renewable energies is one of the main vehicles towards the energy transition, but it can only be realized by extending the hosting capacity of distribution networks. The recent developments on local energy markets, aimed to balance renewable energy flows at the neighborhood scale, have the potential to increase such capacity. Nevertheless, as most scheduling mechanisms, they create further imbalances in flows across phases, which favor higher voltage rises/drops and network losses. 

In this paper, we propose dynamic phase switching as a mechanism to cope with these issues by coordinating phase switching decisions with market decisions. The mechanisms are simple and they rely only on the information obtained from market decisions about the flows exchanged among households. We provide a thorough assessment of the performance of the system, showing that only a small set of households (6\%) need to be equipped with dynamic phase switches to counteract the negative effects of scheduling mechanisms while maintaining all their benefits. Furthermore, the performance obtained when deploying switches in 12\% of total houses is similar to the one observed on a setting without DER. Combining the benefits of local energy markets with cost effective dynamic phase switching mechanisms we can effectively increase the capacity of distribution grid for hosting renewable energy. Furthermore, based on the proposed architecture the QoS can be further improved, by introducing more advanced phase switching decisions, based for example on machine learning, in order to better exploit the data available at DSO.

\bibliographystyle{ACM-Reference-Format}
\bibliography{references} 

\end{document}